\documentclass[12pt,a4paper]{article}

\usepackage{hyperref}

\usepackage{graphicx}
\usepackage{color}

\usepackage{amssymb}
\usepackage{gensymb}
\usepackage{textcomp}
\usepackage[version=4]{mhchem}

\usepackage{setspace}

\title{X-ray absorption and Raman spectroscopy studies of tungstates solid solutions \ce{Zn_cNi_{1-c}WO4} ($c$=0.0-1.0)}

\author{G. Bakradze, A. Kalinko, and A. Kuzmin\\ \\  \small\em Institute of Solid State Physics, University of Latvia,\\ \small\em 8 Kengaraga street, Riga LV-1063, Latvia\\ \small E-mail: georgijs.bakradze@cfi.lu.lv}

\date{Received \today}

\begin{document}
	
\maketitle

\newpage

\begin{abstract}
The influence of thermal disorder and static distortions on the local structure in microcrystalline solid solutions of tungstates \ce{Zn_cNi_{1-c}WO4} with $c$=0.0--1.0  was investigated using temperature-dependent (10--300~K) x-ray absorption spectroscopy (XAS) at the W L$_3$-edge. In addition, the vibrational properties of the solid solutions were studied by Raman spectroscopy. Our results indicate that the formation of solid solutions is accompanied by strong structural relaxations leading to the changes in the \ce{[WO6]} octahedra distortions, which, in turn, affect the vibrational properties of tungstates. In particular, the frequency and band width of the highest W--O stretching mode at about 900~cm$^{-1}$ show distinct composition dependence, following the local structure modifications as revealed by XAS. \\

\end{abstract}

\newpage

\section{Introduction}\label{s:intro}

Versatile physical and chemical properties of tungstates find a wide range of applications in scintillators \cite{Mikhailik2005, Millers1997, Millers1997b,  MOSES2002, Elsts2005, Nikl2008}, down-conversion phosphors \cite{NAGIRNYI2002, Vanetsev2016}, white light-emitting diodes \cite{Zhai2016}, supercapacitors \cite{Niu2013},  lithium-ion batteries \cite{Wang2018} and laser host materials \cite{BASIEV2000, Pask2003}. Recently their use as heterogeneous catalysts \cite{Zhang2014, Yan2019}, humidity or gas sensors \cite{Bhattacharya1997}, electrochromic materials \cite{Kuzmin2001}, anticorrosion pigments \cite{Kalendova2015}, optical temperature sensors \cite{Zhang2018}  and in optical recording \cite{Kuzmin2007, Kuzmin2015}  has been explored too. Many of tungstate functional properties can be further modified by reducing crystallite size, by doping or by making solid solutions with other tungstates \cite{ Grigorjeva2000, Yu2003,   Isupov2005b, Kaczmarek2013, Dey2014, Kuzmin2016}. The solid solution approach is of particular interest due to a wide range of possible chemical compositions.  

A small difference in the size of Ni$^{2+}$ (0.69~\AA) and Zn$^{2+}$ (0.74~\AA) ions \cite{Shannon1976} as well as a small difference in their electronegativity values favour the formation of a continuous series of \ce{Zn_cNi_{1-c}WO4} solid solutions. However, their properties and possible applications were rarely studied in the past. The long-range structure, optical and magnetic properties of \ce{Zn_cNi_{1-c}WO4} solid solutions were characterized in \cite{Oliveira2008, Kalinko2011a, Huang2015}, and their possible use as a yellow pigment has been proposed in \cite{Oliveira2008}.

Pure \ce{NiWO4} and \ce{ZnWO4} belong to the isomorphous series of tungstates with a general composition \ce{MWO4}, where \ce{M^{2+}} is a small-size bivalent cation (with the ionic radius $r_\text{M} < 0.77$ \AA\ \cite{Shannon1976}). They adopt the structural type of wolframite (monoclinic singony, space group $P2/c$ (No. 13)) with the primitive cell containing two formula units ($Z$ = 2) \cite{Keeling1957, Filipenko1968, Trots2009}. Metal ions are octahedrally coordinated by oxygen atoms, and the metal-oxygen octahedra of the same metal type are joined by edges, whereas the octahedra with different metals are connected by corners. The \ce{[MO6]} octahedra are slightly tetragonally distorted, whereas the \ce{[WO6]} octahedra are strongly distorted with tungsten ions being located off-centre due to the second-order Jahn-Teller effect induced by the \ce{W^{6+}}(5d$^0$) electronic configuration  \cite{KUNZ1995}. The distorted octahedra form infinite zigzag chains along the [001] direction; in the perpendicular [100] direction the short chains of \ce{[MO6]} octahedra form a layer which alternates with a layer of edge-joined \ce{[WO6]} octahedra. The lattice dynamics of \ce{NiWO4} and \ce{ZnWO4} have been repeatedly studied using Raman scattering and infrared spectroscopies in \cite{Oliveira2008, Liu1988, Wang1992, Fomichev1994, Kuzmin2001a, Errandonea2008, Kuzmin2011niwo4}.

One should note that below the Ne\'{e}l temperature $T_{\rm N}$ = 67 K the antiferromagnetic ordering occurs in pure \ce{NiWO4} with a doubled magnetic unit cell along the $a$-axis \cite{WEITZEL1970}. The local magnetic interactions are determined by the spins at \ce{Ni^{2+}} ions, being coupled ferromagnetically in the same chain of the \ce{[NiO6]} octahedra, but antiferromagnetically in adjacent chains \cite{Kuzmin2011niwo4}. At the same time, \ce{ZnWO4} is a wide band gap insulator ($E_\text{g} \approx 4.6$ eV \cite{Kalinko2009}). One can expect that a dilution of nickel ions by non-magnetic zinc ions in \ce{Zn_cNi_{1-c}WO4} solid solutions will affect the magnetic ordering and modify local interactions, which, in turn, should influence the local structure and lattice dynamics of the compound. 

In this study, we have used the temperature-dependent (10--300~K) W L$_3$-edge x-ray absorption spectroscopy (XAS) to probe the local structure in a series of \ce{Zn_cNi_{1-c}WO4} solid solutions and Raman scattering spectroscopy to follow composition dependence of their vibrational properties.

\section{Experimental and data analysis}\label{s:exper}
\ce{Zn_cNi_{1-c}WO4} solid solutions were co-precipitated at room temperature (20~$\celsius$) and pH=8 using \ce{ZnSO4.7H2O}, \ce{Ni(NO3)2.6H2O}, and \ce{Na2WO4.H2O} as educts \cite{Kalinko2011a}. First, double distilled water solutions of the three salts were prepared. Next, the nitrate solutions were mixed in the stoichiometric ratios and added to the tungstate solution to obtain \ce{Zn_cNi_{1-c}WO4} with $c$=0.0--1.0. The products precipitated immediately upon mixing the aqueous solutions. After completion of the reaction, the precipitates were filtered off, washed several times with double distilled water, and dried in air for 12 hours at 80~$\celsius$. Microcrystalline \ce{Zn_cNi_{1-c}WO4} powders were produced by annealing the precipitates in ambient atmosphere for 4 hours at 800~$\celsius$. The phase composition of the samples was controlled by x-ray powder diffraction (XRD) \cite{Kalinko2011a}.

Raman scattering spectra were collected at room temperature using a confocal microscope with a spectrometer Nanofinder-S (SOLAR TII, Ltd.) \cite{Kuzmin2007}. To excite the Raman spectra, we used a He–Cd laser radiation (441.6~nm, 50~mW cw power) dispersed by an 1800~grooves/mm diffraction grating  mounted in the 520~mm focal length monochromator. The measurements were performed through an optical objective (Nikon Plan Fluor 40$\times$, NA=0.75) using a Peltier-cooled back-thinned CCD camera (ProScan HS-101H) as a detector. An edge filter (Omega, 441.6AELP-GP) was used to eliminate the elastic laser light components with the cut-off wavenumber around 270~cm$^{-1}$. All spectra were normalized to the intensity of the strongest Raman band at about 900~cm$^{-1}$.

X-ray absorption measurements were performed in transmission mode at the HASYLAB DESY C1 bending-magnet beamline \cite{DORISC} in the temperature range from 10~K to 300~K at the W L$_3$ (10207~eV) edge. The storage ring DORIS III operated at $E$=4.44~GeV and $I_\text{max}$=140~mA. The x-ray beam intensity was measured by two ionization chambers filled with argon and krypton gases. The higher-order harmonics were effectively eliminated by detuning the double-crystal monochromator Si(111) to 60\% of the rocking curve maximum, using the beam-stabilization feedback control. The Oxford Instruments LHe flow through cryostat was used to maintain the required sample temperature. The powder samples were deposited on Millipore filters and fixed by Scotch tape. The deposited powder weight was chosen to give the value of the absorption edge jump close to 1.0.

The XAS spectra were analysed using the XAESA software package \cite{XAESA}. The W L$_3$ extended x-ray absorption fine structure (EXAFS) $\chi(k)k^2$ was extracted from the total absorption spectra using conventional procedure \cite{Kuzmin2014} as a function of the photoelectron wavenumber $k = \sqrt{(2 m_\text{e}/\hbar^2) (E - E_0)}$, where $E$ is the photon energy, $E_0$ is the minimal energy to excite a core electron from the W L$_3$ shell, $m_\text{e}$ is the electron mass, and $\hbar$ is the reduced Planck’s constant. The origin of the photoelectron energy $E_0$ was set at 10210.5~eV for the W L$_3$-edge to have the best match between the experimental and calculated EXAFS spectra. To determine the true radial distribution functions (RDFs) from the EXAFS spectra, a regularization-like method \cite{Kuzmin2000, ANSPOKS2014} was used. The scattering amplitude and phase shift functions for the W--O pairs required for the simulation were calculated by the ab initio multiple-scattering FEFF8.5L code using the complex exchange-correlation Hedin-Lundqvist potential \cite{Ankudinov1998, Rehr2000}. The calculations were performed based on the crystallographic structure of \ce{ZnWO4} \cite{Filipenko1968} or \ce{NiWO4} \cite{Keeling1957}, considering an 8~\AA\  cluster around the absorbing tungsten atom. Calculations of the cluster potentials were done in the muffin-tin (MT) self-consistent-field approximation using the default values of the MT radii as implemented in the FEFF8.5L code \cite{Ankudinov1998}. Only the first coordination shell contributions, singled out by the back-Fourier transformation procedure in the range of $R$=0.75--2.85~\AA, were considered in the analysis. The best-fits of EXAFS for the first coordination shell of tungsten were performed in the $k$-space range of 2.0--15.0~\AA$^{-1}$.

\section{Results and discussion}\label{s:results}

\subsection{Raman scattering spectra}\label{ss:raman}
\begin{figure}[t]
	\centering
	\includegraphics[width=0.65\textwidth]{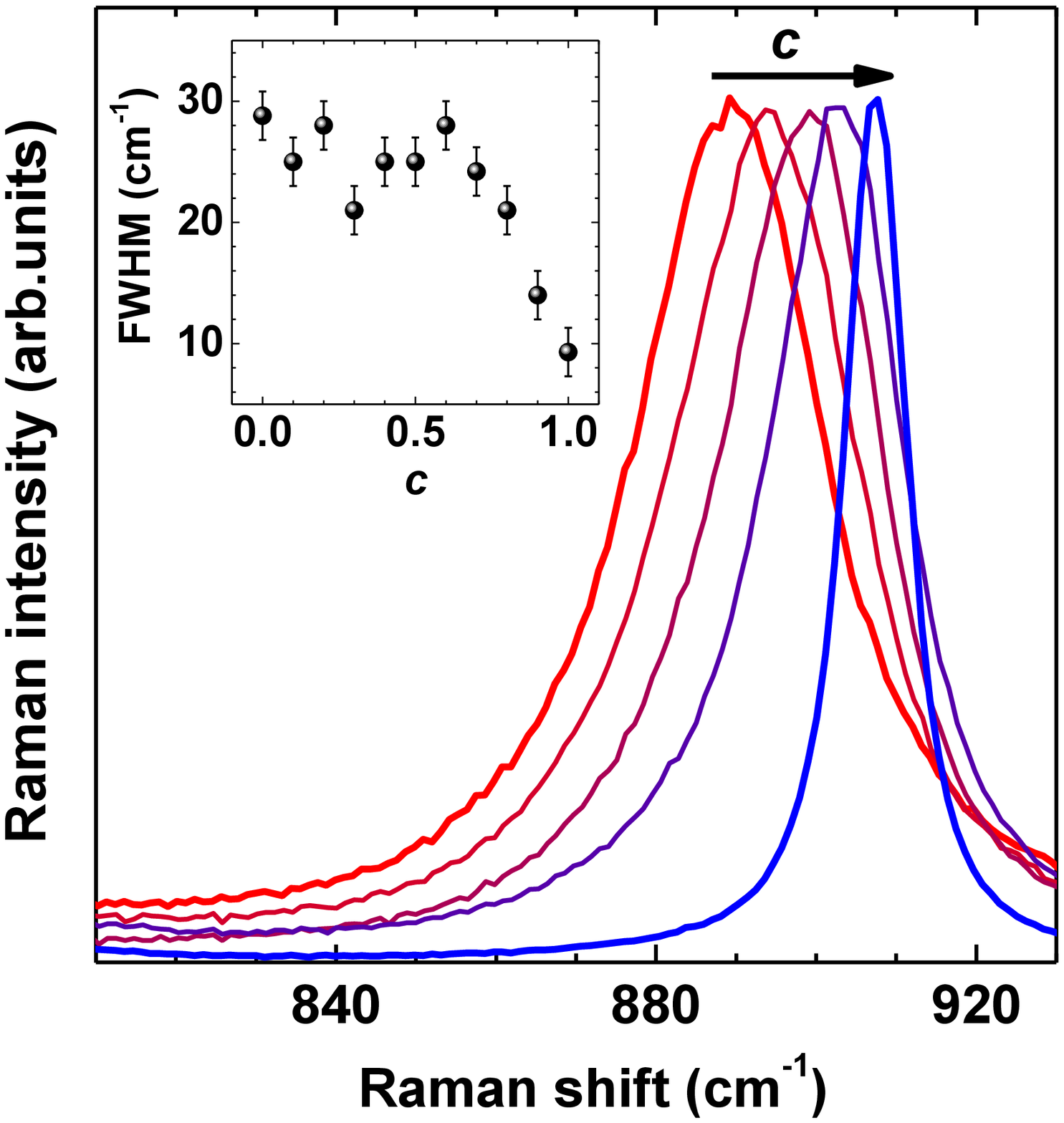}
	\caption{Raman scattering spectra of \ce{Zn_cNi_{1-c}WO4} solid solutions in the range of the $\mathrm{A_g}$ band due to the W--O stretching mode. For the sake of clarity only the Raman spectra for $c$ = 0.0, 0.2, 0.4, 0.8, 1.0 are shown. The inset shows FWHM of the $\mathrm{A_g}$ band as a function of composition.}
	\label{fig1}
\end{figure}

Figure~\ref{fig1} shows the Raman scattering spectra for selected \ce{Zn_cNi_{1-c}WO4} solid solutions in the $\mathrm{A_g}$ band region around 900~cm$^{-1}$  \cite{Liu1988,Wang1992} and the dependence of the $\mathrm{A_g}$ band  position on the Zn content $c$. The $\mathrm{A_g}$  band has the highest intensity among all bands and is due to the symmetric stretching W--O vibrations in the \ce{[WO6]} octahedral oxyanion groups \cite{Liu1988, Wang1992, Kuzmin2011niwo4,  Kalinko2009lum}. Therefore, the $\mathrm{A_g}$ band should be sensitive to the \ce{[WO6]} octahedra distortion and W--O bond strength \cite{Daniel1987}.  

\begin{figure}[t]
	\centering
	\includegraphics[width=0.65\textwidth]{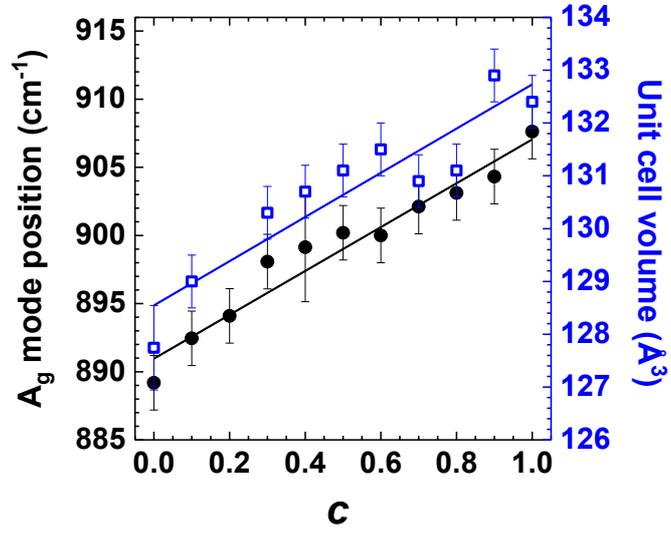}
	\caption{Dependence of the $\mathrm{A_g}$ band position (filled symbols) and the unit cell volume (hollow symbols) in the series of \ce{Zn_cNi_{1-c}WO4} solid solutions on Zn content $c$. The lines are the least-square fits obeying $\omega_\text{W-O}(c) = (891 \pm 1) + (16 \pm 1) c$ and $V(c) = (128.5 \pm 0.4) + (4.2 \pm 0.6) c$.}
	\label{fig2}
\end{figure}

The position of the $\mathrm{A_g}$ band shifts monotonously (Fig.~\ref{fig2}) with increasing the degree of Ni substitution by Zn from 889~cm$^{-1}$ in \ce{NiWO4} ($c = 0$) to 908~cm$^{-1}$ in \ce{ZnWO4} ($c = 1$). At the same time, the unit cell volume of \ce{Zn_cNi_{1-c}WO4} solid solutions, taken from \cite{Kalinko2011a}, increases going from \ce{NiWO4} to \ce{ZnWO4} that suggests the opposite dependence of the $\mathrm{A_g}$ stretching mode frequency on the zinc content. Note that the width of the $\mathrm{A_g}$ band is larger for the low zinc content: the full width at half maximum (FWHM) is $\approx 25$~cm$^{-1}$ for $c < 0.8$ but decreases down to $\approx 10$~cm$^{-1}$ for $c = 1$. Such behaviour of the  $\mathrm{A_g}$ band in the Raman scattering spectra of \ce{Zn_cNi_{1-c}WO4} solid solutions can be explained by the interplay between Ni--O, Zn--O, and W--O bond strengths. More covalent Ni--O bonds,  compared to the Zn--O bonds, attract electrons that leads to a weakening of the W--O bond and, consequently, to a lower frequency and broadening of the $\mathrm{A_g}$ band in \ce{NiWO4}.

\subsection{X-ray absorption spectroscopy}\label{ss:xas} 
\begin{figure}[t]
	\centering
	\includegraphics[width=1\textwidth]{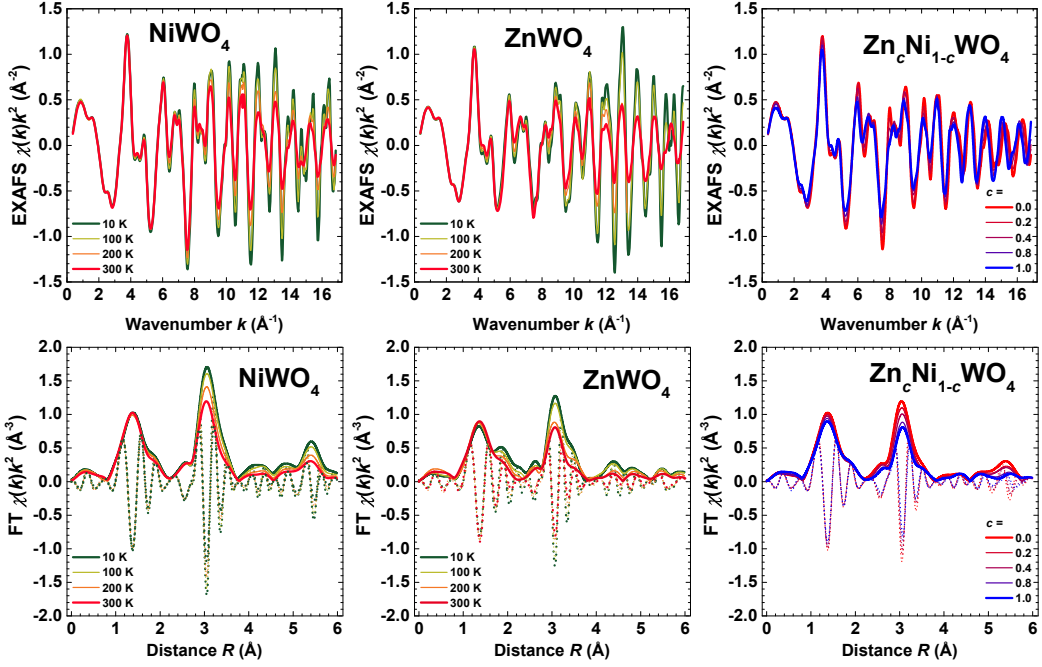}
	\caption{W L$_3$-edge EXAFS spectra $\chi(k)k^2$ and their Fourier transforms (FTs) for microcrystalline \ce{NiWO4}, \ce{ZnWO4}, and \ce{Zn_cNi_{1-c}WO4} solid solutions. Only the modulus and imaginary part are shown in FTs. The data are presented as a function of temperature and Zn content $c$ for pure components and \ce{Zn_cNi_{1-c}WO4} solid solutions, respectively. Note that the positions $R$ of the peaks in FTs differ from the crystallographic values (see text for details) due to the phase shift present in EXAFS.}
	\label{fig3}
\end{figure}

Temperature-dependent (10--300 K) experimental W L$_3$-edge EXAFS spectra $\chi(k)k^2$ and their Fourier transforms (FTs) for microcrystalline \ce{NiWO4} and \ce{ZnWO4} are shown in Fig.~\ref{fig3}. The first broad peak in the FTs at $R$=0.8--2.2~\AA\ corresponds to the first coordination shell of tungsten, whereas the strong peak at $R \simeq 3$~\AA\ is mainly due to metal atoms in the second coordination shell \cite{timoshenko2014cuwo4, timoshenko2015cowo4, timoshenko2015mncowo4}. Here we focus on the analysis of the first coordination shell only, however, the structural peaks in FTs are well resolved at least till 6~\AA. 

In the W L$_3$-edge EXAFS spectra, the effect of thermal disorder is well apparent and leads to the EXAFS oscillations damping at large $k$-values (Fig.~\ref{fig3}). Thermal disorder strongly affects the amplitude of the peaks in FTs due to the longest W--O bonds in \ce{[WO6]} octahedra (the shoulder at $R \approx 1.9 $~\AA) and outer coordination shells ($R > 2$~\AA). At the same time, the contribution from the group of nearest four oxygen atoms, responsible for the peak in FTs at $R = 1.4$~\AA, is almost temperature independent. This finding testifies to a strong interaction between tungsten and nearest oxygen atoms within the oxyanion groups.  

Changes in the composition have a clear effect on the local structure of \ce{Zn_cNi_{1-c}WO4} solid solutions: the W L$_3$-edge  EXAFS shape in the whole $k$-range, as well as the amplitude of all peaks in FTs are affected (Fig.~\ref{fig3}). 

\begin{figure}[t]
	\centering
	\includegraphics[width=1\textwidth]{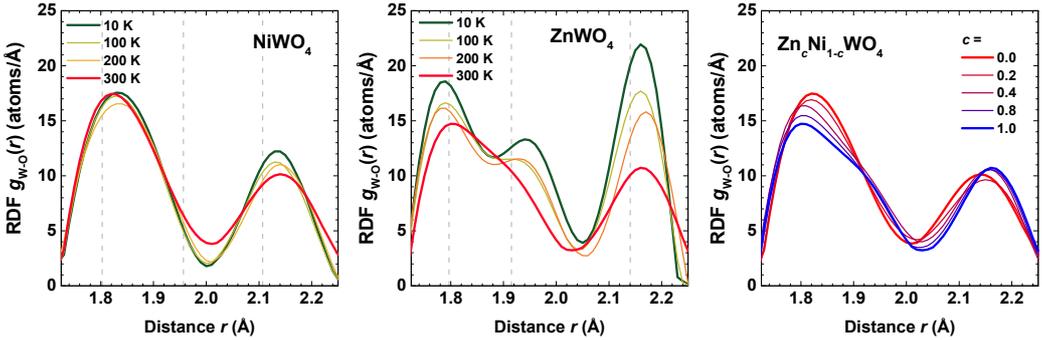}
	\caption{Reconstructed RDFs $g_{\rm W-O}(r)$ for the first coordination shell of  tungsten in microcrystalline \ce{NiWO4} and \ce{ZnWO4} as a function of temperature in the range of 10--300 K, and in microcrystalline \ce{Zn_cNi_{1-c}WO4} at 300~K as a function of zinc content $c$. The gray vertical lines indicate the W--O bond lengths as-determined from XRD (see text for details).}
	\label{fig4}
\end{figure}

In pure tungstates, the \ce{[WO6]} octahedra are strongly distorted due to the second-order Jahn-Teller effect \cite{KUNZ1995}. There are  three pairs of non-equivalent W--O bonds: 2$\times$1.803~\AA, 2$\times$1.957~\AA, and 2$\times$2.107~\AA\ in \ce{NiWO4} \cite{Keeling1957} and 2$\times$1.797~\AA, 2$\times$1.915~\AA, and 2$\times$2.140~\AA\ in \ce{ZnWO4} \cite{Filipenko1968}. Such distortions are responsible for the splitting of the W--O radial distribution functions (RDFs) $g_{\rm W-O}(r)$ within the first coordination shell of tungsten into subshells, being well distinguishable in Fig.~\ref{fig4}. In \ce{NiWO4}, the RDFs exhibit two broad peaks with a weak temperature dependence (especially for the distant groups of oxygen atoms located at 2.0--2.3~\AA). At the same time, the RDFs in \ce{ZnWO4} have three-peak shape and are more sensitive to thermal disorder.

In \ce{Zn_cNi_{1-c}WO4} solid solutions at room temperature (Fig.~\ref{fig4}), the reconstructed RDFs $g_{\rm W-O}(r)$ show a distinct composition dependence, and their shapes are framed by the two RDFs of pure components. The behaviour of the nearest group of O atoms around W upon Ni substitution with Zn can be correlated with that of the $\mathrm{A_g}$ mode in the Raman scattering spectra (cf. Fig.~\ref{fig1}). An increase of the W--O stretching mode frequency for the Zn-rich samples suggests the presence of stronger (and shorter) bonds between W and nearest O atoms, which manifests itself in RDFs as a small shift in the maximum of the first peak in RDFs from $r = $1.82~\AA\ to 1.80~\AA\ (whereas the bonds to the distant O atoms become longer, i.e. less strongly bound, cf. Fig.~\ref{fig4}). At the same time, an increase of the W--O stretching mode frequency for samples with a higher Zn content indicates weakening of more covalent Ni--O bonds, being consistent with the presence of the antiferromagnetic exchange coupling in pure \ce{NiWO4} \cite{WEITZEL1970, Anderson1959}.

\section{Conclusions}\label{s:conc}

The influence of thermal disorder and static distortions on the local structure in microcrystalline \ce{Zn_cNi_{1-c}WO4} solid solutions with $c$=0.0--1.0 was studied by Raman scattering spectroscopy and temperature-dependent (10--300~K) x-ray absorption spectroscopy at the W L$_3$-edge. 

The room-temperature Raman scattering spectra of \ce{Zn_cNi_{1-c}WO4} solid solutions demonstrate strong composition dependence. The frequency of the most intense $\mathrm{A_g}$ band -- due to the W--O stretching mode -- increases by about 19~cm$^{-1}$ from \ce{NiWO4} to \ce{ZnWO4}, whereas its full width at half maximum decreases by about 15~cm$^{-1}$ for $c > 0.8$. The observed changes in the vibrational dynamics are caused by the interplay between Ni--O, Zn--O, and W--O interaction, leading to the change in local distortions of the oxyanion groups \ce{[WO6]} upon Ni substitution with Zn in \ce{Zn_cNi_{1-c}WO4}. 

Analysis of temperature-dependent W L$_3$-edge extended x-ray absorption fine structure spectra made it possible to separate the effects of dynamic disorder and static distortions in the first coordination shell of the W atoms. In \ce{NiWO4}, the static distortions dominate while thermal effects mainly influence the distant group of O atoms at about 2.0--2.3~\AA. The deformation of the \ce{[WO6]} octahedra is more pronounced in \ce{ZnWO4} than in \ce{NiWO4}, where three groups of the W--O bonds can be identified at 1.78~\AA, 1.95~\AA, and 2.15~\AA\ at low temperatures.

The composition-induced distortions of the \ce{[WO6]} octahedra in \ce{Zn_cNi_{1-c}WO4} solid solutions were evidenced from the shape of the W--O radial distribution functions. With increasing Zn content, the lengths of the nearest W--O bonds become shorter, while the lengths of the distant W--O bonds become longer. This observation correlates well with the increase of the W--O stretching mode frequency in the Raman scattering spectra for the larger $c$. These results can be explained by the competing effect on W--O bonds of two  bonds: weaker Zn--O and stronger Ni--O.

\section*{Acknowledgements}

G.B. acknowledges the financial support provided by the State Education Development Agency for project No. 1.1.1.2/VIAA/3/19/444 (agreement No. 1.1.1.2/16/I/001) realized at the Institute of Solid State Physics, University of Latvia. A.K. and A.K. would like to thank the support of the Latvian Council of Science project No. lzp-2019/1-0071. Institute of Solid State Physics, University of Latvia as the Center of Excellence has received funding from the European Union's Horizon 2020 Framework Programme H2020-WIDESPREAD-01-2016-2017-TeamingPhase2 under grant agreement No. 739508, project CAMART2.

\bibliography{references}

\newpage

\newpage 

\newpage 

\newpage 

\end{document}